\documentclass[conference]{IEEEtran}
\IEEEoverridecommandlockouts
\usepackage{cite}
\usepackage{amsmath,amssymb,amsfonts}
\usepackage{algpseudocode}
\usepackage{algorithm}
\usepackage{subcaption}
\usepackage{graphicx}
\usepackage{graphicx}
\usepackage{textcomp}
\usepackage{xcolor}
\usepackage{siunitx}
\usepackage{booktabs}
\usepackage{multirow}
\usepackage{comment}
\usepackage{bm}
\usepackage{hyperref}
\def\BibTeX{{\rm B\kern-.05em{\sc i\kern-.025em b}\kern-.08em
    T\kern-.1667em\lower.7ex\hbox{E}\kern-.125emX}}
\begin{document}

% \title{
% A Self-Attention Based Pipeline for Unit Commitment
%}
\title{
A Multi-Stage Warm-Start Deep Learning Framework for Unit Commitment
% Feasibility-Enhanced Warm-Start for Multi-Day Unit Commitment
}

\author{
\IEEEauthorblockN{Muhy Eddin Za'ter, Anna Van Boven, Bri-Mathias Hodge and Kyri Baker}
\IEEEauthorblockA{\textit{University of Colorado Boulder} \\
Boulder, CO, USA \\
\{muhy.zater, anna.vanboven, brimathias.hodge, kyri\}@colorado.edu}
}

\maketitle

\begin{abstract}
Maintaining instantaneous balance between electricity supply and demand is critical for reliability and grid instability. System operators achieve this through solving the task of Unit Commitment (UC),ca high dimensional large-scale Mixed-integer Linear Programming (MILP) problem that is strictly and heavily governed by the grid physical constraints. As grid integrate variable renewable sources, and new technologies such as long duration storage in the grid, UC must be optimally solved for multi-day horizons and potentially  with greater frequency. Therefore, traditional MILP solvers increasingly struggle to compute solutions within these tightening operational time limits. To bypass these computational bottlenecks, this paper proposes a novel framework utilizing a transformer-based architecture to predict generator commitment schedules over a 72-hour horizon. Also, because raw predictions in highly dimensional spaces often yield physically infeasible results, the pipeline integrates the self-attention network with deterministic post-processing heuristics that systematically enforce minimum up/down times and minimize excess capacity. Finally, these refined predictions are utilized as a warm start for a downstream MILP solver, while employing a confidence-based variable fixation strategy to drastically reduce the combinatorial search space. Validated on a single-bus test system, the complete multi-stage pipeline achieves 100\% feasibility and significantly accelerates computation times. Notably, in approximately 20\% of test instances, the proposed model reached a feasible operational schedule with a lower overall system cost than relying solely on the solver.
\end{abstract}

\begin{IEEEkeywords}
Unit commitment, learn to optimize, Deep Learning.
\end{IEEEkeywords}

\section{Introduction}

% Done
Maintaining instantaneous balance between electricity supply and demand is critical to prevent grid instability and cascading blackouts\cite{grigsby2007power}. Independent System Operators (ISOs) achieve this by continuously clearing wholesale electricity markets through Unit Commitment (UC)\cite{giannakis2013monitoring}, an optimization challenge that determines the optimal binary startup and shutdown sequences of generators over a scheduled time horizon. Because operational decisions are temporally coupled and governed by strict physical constraints, UC is formulated as a large-scale Mixed-Integer Linear Programming (MILP) problem\cite{abdou2018unit}. The computational complexity of UC scales exponentially with the number of generators, making it an NP-hard problem\cite{bendotti2019complexity}.

This bottleneck is further exacerbated by grid decarbonization goals. This transition toward variable and uncertain electricity sources such as solar and wind introduces complex stochastic variables and requires more restrictive models to compensate for the decreasing rotational inertia of the grid\cite{bahrami2022neural}. Consequently, traditional MILP solvers increasingly struggle to compute solutions within the strict operational time limits required by ISOs \cite{feas_restore}. Furthermore, as grid planning evolves to evaluate the impact of long-duration energy storage and multi-day weather events, operators must increasingly expand the UC horizon from the standard 24-hour day-ahead markets to longer periods\cite{thatte2024role}. All of the above-mentioned aspects necessitate research efforts toward alternative solutions toward faster UC. Because UC is solved repeatedly with identical structural topologies but varying exogenous data inputs, Machine Learning (ML) has emerged as a potential new transformative paradigm to bypass these computational limits.

While ML offers significant speedups by predicting generator on/off statuses, thereby transforming the computationally expensive MILP into a faster Linear Program (LP), minor statistical errors in this highly dimensional space easily yield physically infeasible results\cite{yang2021machine, ramesh2023feasibility}. To bridge this gap, the literature proposes various methods \cite{aiuc_lr}, including entirely bypassing the MIP solver via specific architectures \cite{gan, gcn_onoff, dl_gru}, or reducing MILP dimensionality by predicting statuses only for generators with high prediction confidence\cite{milp_lp_short},  utilizing predictions as solver warm-starts \cite{rl_warmstart}, or introducing deterministic post-processing heuristics to restore feasibility \cite{feas_restore}. However, a critical limitation remains: traditional neural architectures struggle to process extremely high-dimensional state spaces while effectively capturing complex temporal couplings, such as multi-hour ramping constraints and minimum up/down times\cite{nguyen2025fsnet}. This shortfall is highly evident when expanding to longer horizons, potentially leading to sub-optimality, infeasibility, and reduced computational gains.

To address these limitations, we propose a novel framework that employs a transformer-based architecture to natively learn deep temporal dependencies over a multi-day ahead horizon, and utilizes the most effective ML-based strategies from the recent literature\cite{yang2021machine, ramesh2023feasibility, nguyen2025fsnet}. By synergizing our self-attention base sequence-modeling network \cite{vaswani2017attention} with post-processing heuristics, predictive warm-start, and targeted variable fixation, our pipeline bridges the gap between raw statistical predictions and the strict physical constraints of the UC problem over a long horizon.

The contributions of this paper are as follows:

\begin{itemize}
    \item \textbf{Self-Attention for Temporal Coupling:} We introduce an architecture that leverages multi-head self-attention to explicitly model the long-range temporal dependencies and high-dimensional interactions of the grid over a multi-day-ahead horizon. We also utilize zero-initialized learnable positional embeddings, where the model inherently captures the periodic structure of daily load and renewable generation curves.

\item \textbf{Multi-Stage Feasibility Pipeline:} We detail three deterministic post-processing algorithms that systematically enforce minimum up/down time constraints, minimize excess online capacity via localized economic dispatch, and optimize generator utilization.

\item \textbf{Confidence-based Search Space Reduction:} We implement a targeted variable fixation strategy that sets generator statuses falling above or below specific predictive confidence thresholds when the models predictions are passed as warm starts. This approach reduces the combinatorial dimensionality of the downstream MILP solver, accelerating computation times by leaving only the most marginal decisions to for the solver.

\item We open source the data, data generation pipeline and the source code of the proposed method in this repository\footnote{https://github.com/CUgriffinlab/AI4UnitCommitment}.

\end{itemize}

% The rest of the paper is organized as follows. Section II presents in detail the methodology, including the model, the heursitics based post-processing and warm start of the solver. Section III shows the implementation details, validation and results of the conducted experiments.
\section{Methodology}\label{sec: method}

\begin{figure*}[!h]
\label{fig:pipeline}
    \centering
    \includegraphics[width=0.65\linewidth 
    ]{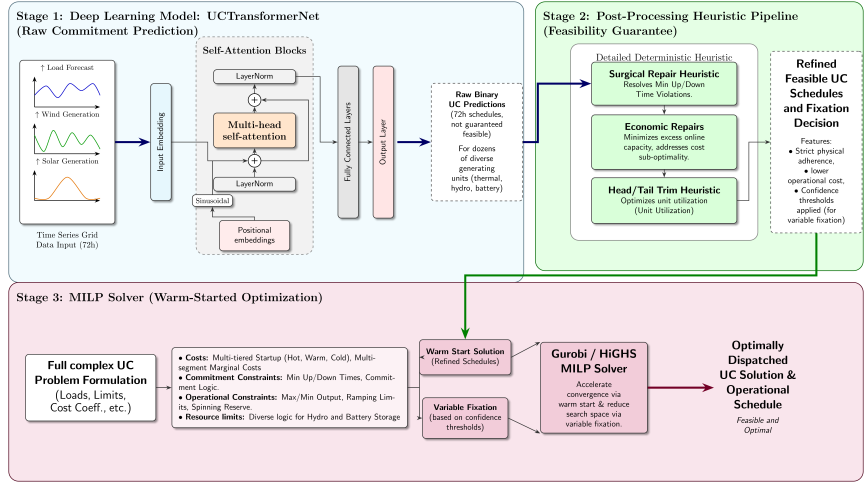}
    \caption{The 3 stages of the proposed pipeline. Stage 1 is the deep learning model. Stage 2 is the post-processing heuristics to restore feasibility. Stage 3 is the warm-started MILP solver.}
    \label{fig:placeholder}
\end{figure*}

As discussed above, machine learning has the potential to drastically speed up obtaining solutions to large optimization problems, where computation time is currently a bottleneck. However, these ML predictions often cannot be relied on independently unless they can provide feasibility and optimality guarantees. In this section, we introduce the baseline unit commitment problem we seek to solve (\ref{sec:prob_form}), then discuss the model used to predict generator statuses (\ref{sec:model}) and the data generated to train the model (\ref{sec:data}). The post-processing steps used to improve the model prediction are then described in subsections \ref{sec:heur} and \ref{sec:ws}. Figure 1 shows the overall pipeline of the proposed method.

\subsection{Unit Commitment for a Single-Bus Test System}\label{sec:prob_form}

This paper presents a model submitted to EPRI's `AI-ccelerating Unit Commitment' competition, a four-month competition organized to foster research in ML-enhanced algorithm development for solving the unit commitment problem \cite{epri}. The model presented here finished fourth out of over 2,300 submissions from a total of 123 teams participating. 

The model was tasked with predicting generator on-off statuses for a single-bus test system comprised of 51 thermal units, 15 hydro plants, 13 batteries and 1 pumped hydro storage. The system also has 1 source each of aggregated solar power, aggregated wind power, and aggregated system load (perfect forecasting is assumed for each of these sources). The competition's objective is to determine the generator statuses for each generator over a 72-hour period that minimizes the system cost. The  predictions need to be feasible; there must be sufficient generation to meet demand, and each generator must not violate start-up/shut-down or ramp-up/down constraints. All 51 generators are commitable generators.

\subsection{Model}\label{sec:model}

\begin{figure*}[!h]
    \centering
    \includegraphics[width=0.55\linewidth
    ]{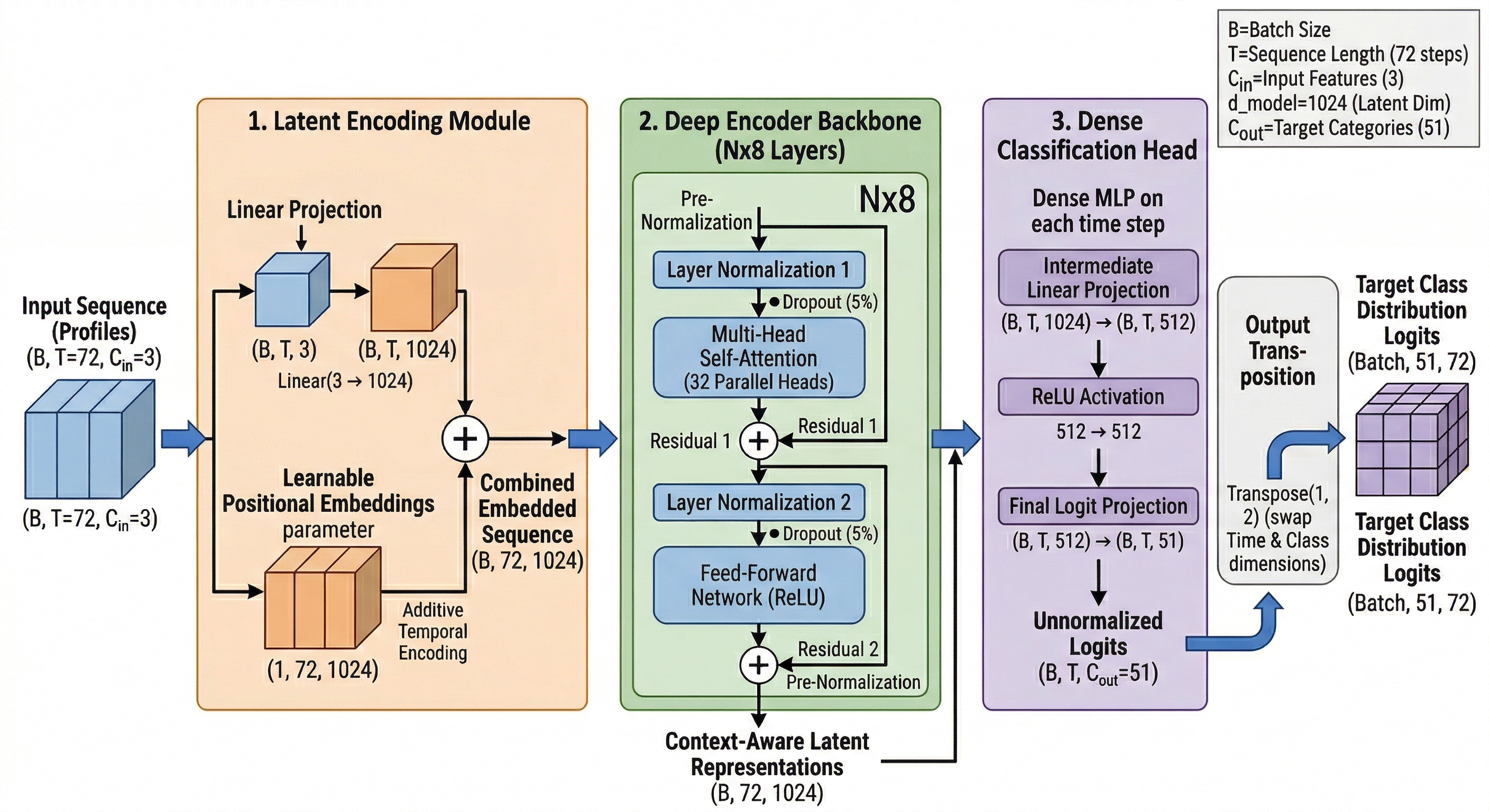}
    \caption{The components of the proposed architecture. The latent component is used to process the input with positional encoding. The deep encoder consists of self-attention to map the input to the output. Finally, the classification head is responsible for generating the final predictions.}
    \label{fig:arch}
\end{figure*}

We propose a transformer-based neural network designed to predict generator commitment schedules for a 72-hour planning horizon, given the load, solar, and wind profiles. The model outputs binary commitment decisions for all $N_g$ thermal generators simultaneously, which are subsequently refined by a feasibility post-processing heuristic before being passed to an economic dispatch LP solver. The architecture of the model is visualized in Fig. \ref{fig:arch}.

\subsubsection{Problem Formulation}
\label{subsec:formulation}
Let $\mathbf{P} \in \mathbb{R}^{B \times T \times F}$ denote a batch of $B$ input profiles over $T = 72$ hourly time steps, where each step contains $F = 3$ features: normalized system demand, solar generation, and wind generation. The raw profiles are normalized using training-set statistics ($\bm{\mu}, \bm{\sigma}$) to stabilize gradient flow across inputs with different physical scales:
{\small
\begin{equation}
    \hat{\mathbf{p}}_t = \frac{\mathbf{p}_t - \bm{\mu}}{\bm{\sigma} + \epsilon}, \quad \epsilon = 10^{-6}
\end{equation}
The model produces a logit tensor $\mathbf{Z} \in \mathbb{R}^{B \times N_g \times T}$, from which binary commitment decisions are recovered via a sigmoid function and a threshold $\tau$ tuned on validation instances:
{\small
\begin{equation}
    u_{i,t} = \mathbf{1}\!\left[\sigma(z_{i,t}) > \tau\right], \quad i = 1, \ldots, N_g, \quad t = 1, \ldots, T
\end{equation}}

\subsubsection{Input Encoding}
\label{subsec:encoding}
The input sequence $\hat{\mathbf{P}}$ is first projected into the model's latent dimension $d_{\text{model}}$ via a learned linear embedding. To inject temporal position information, a fully learnable positional embedding $\mathbf{\Phi} \in \mathbb{R}^{1 \times T \times d_{\text{model}}}$ is added element-wise:
{\small
\begin{equation}
    \mathbf{H}^{(0)} = \left(\hat{\mathbf{P}}\,\mathbf{W}_e + \mathbf{b}_e\right) + \mathbf{\Phi}, \quad \mathbf{\Phi} \sim \mathcal{N}(0, 0)
\end{equation}}

Unlike standard sinusoidal embeddings, initializing $\mathbf{\Phi}$ to zero allows the model to learn the specific daily periodic structure of grid load and renewable curves entirely from scratch without imposing rigid priors.

\subsubsection{Transformer Encoder}
\label{subsec:transformer}
The encoded sequence $\mathbf{H}^{(0)}$ is processed by $L = 8$ stacked Transformer encoder layers \cite{vaswani2017attention}. Instead of standard post-normalization, each layer applies Pre-Layer Normalization (Pre-LN) \cite{xiong2020layer} to prevent vanishing gradients during early training. 

Within each layer, a Multi-Head Self-Attention (MHSA) module explicitly models the temporal coupling across the 72-hour window. For a power systems context, this mechanism dynamically evaluates how the state of the grid at one specific hour (the Query) is constrained by conditions at all other hours (the Keys), allowing the network to natively learn complex inter-hour constraints like prolonged ramping trajectories. The attention operation for each head $j$ is defined as:
{\small
\begin{align}
    \text{Attention}(\mathbf{Q}_j, \mathbf{K}_j, \mathbf{V}_j) &= \text{softmax}\!\left( \frac{\mathbf{Q}_j\mathbf{K}_j^\top}{\sqrt{d_k}} \right)\mathbf{V}_j \\
    \text{MHSA}(\mathbf{H}) &= \left[ \text{head}_1 \,\|\, \cdots \,\|\, \text{head}_h \right]\mathbf{W}_O
\end{align}
}
where $\mathbf{Q}_j, \mathbf{K}_j, \mathbf{V}_j$ are linear projections of the layer input into a reduced dimension $d_k = d_{\text{model}} / h$, using $h=64$ parallel heads. The MHSA output is then added via a residual connection:
{\small
\begin{equation}
    \tilde{\mathbf{H}}^{(\ell)} = \mathbf{H}^{(\ell-1)} + \text{MHSA}\!\left(\text{LN}\!\left(\mathbf{H}^{(\ell-1)}\right)\right)
\end{equation}
}
\begin{comment}
    
\begin{align}
    \text{Attention}(\mathbf{Q}_j, \mathbf{K}_j, \mathbf{V}_j) &= \text{softmax}\!\left( \frac{\mathbf{Q}_j\mathbf{K}_j^\top}{\sqrt{d_k}} \right)\mathbf{V}_j \\
    \text{MHSA}(\mathbf{H}) &= \left[ \text{head}_1 \,\|\, \cdots \,\|\, \text{head}_h \right]\mathbf{W}_O
\end{align}
where $\mathbf{Q}_j, \mathbf{K}_j, \mathbf{V}_j$ are linear projections of the layer input into a reduced dimension $d_k = d_{\text{model}} / h$, using $h=64$ parallel heads. The MHSA output is then added via a residual connection:
\begin{equation}
    \tilde{\mathbf{H}}^{(\ell)} = \mathbf{H}^{(\ell-1)} + \text{MHSA}\!\left(\text{LN}\!\left(\mathbf{H}^{(\ell-1)}\right)\right)
\end{equation}
\end{comment}

Finally, the sequence is refined by a standard position-wise Feed-Forward Network (FFFN) with dropout ($p = 0.05$) for regularization.

\subsubsection{Commitment Classifier}
\label{subsec:classifier}
To map the abstract temporal representations back to discrete physical unit decisions, the final encoder output $\mathbf{H}^{(L)}$ is passed through a two-layer generator classification head (inner dimension $1024$, ReLU activation). This classification head projects the latent sequence into the non-normalized commitment logits for all $N_g$ thermal generators simultaneously. The resulting tensor is transposed to yield the standard UC format $\mathbf{Z} \in \mathbb{R}^{B \times N_g \times T}$, ensuring the downstream post-processing pipeline receives a cohesive 72-hour schedule for every unit.

\subsubsection{Training Objective}
\label{subsec:training}
Let $\mathbf{Y} \in \{0, 1\}^{B \times N_g \times T}$ denote the ground-truth MILP commitment schedule. The model is trained with a class-weighted binary cross-entropy (BCE) loss\cite{zhou2021machine}. In grid operations, failing to commit a necessary generator (under-commitment) causes physical violations and infeasibility, whereas turning on an extra generator (over-commitment) merely incurs an economic penalty. Therefore, we introduce a positive-class weight $\alpha = 1.5$ to explicitly penalize under-commitment errors:
{\small
\begin{equation}
\begin{split}
    \mathcal{L} = -\frac{1}{B \cdot N_g \cdot T} \sum_{b,i,t} \Big[ &\alpha \cdot y_{b,i,t} \log \sigma(z_{b,i,t}) \\
    &+ (1 - y_{b,i,t}) \log\!\left(1 - \sigma(z_{b,i,t})\right) \Big]
\end{split}
\end{equation}}
The network is optimized using AdamW \cite{loshchilov2019decoupled} with weight decay ($\lambda = 10^{-4}$) and a One-Cycle learning rate schedule \cite{smith2019super}.
 
% ------------------------------------------------------------------
Implementation details and code are available on this repository\footnote{https://github.com/CUgriffinlab/AI4UnitCommitment}.

\subsection{Refinements}\label{sec:heur}
The model initially predicts a  value from $0$ to $1$ for each generator at each hour indicating the probability the generator is on at time $T$. The model uses a threshold of $0.5$ to map predictions to either 0 (off) or 1 (on). This subsection details the three post-processing steps that are used to manually adjust the model's initial prediction. The remainder of this subsection will use the following notation. 
Let $g_i \in G, 1 \leq i \leq 51$ be the set of thermal generators. 
Each generator has a cost term $c_i$, a minimum up ($mu_i$) and minimum down ($md_i$) time, a ramp up ($ru_i$) and ramp down ($rd_i$) limit, and a minimum and maximum generation ($\underline{p_i}, \text{  } \overline{p_i}$). 
For each timestep $t, 1 \leq t \leq 72$, define the on/off status of generator $i$ as $s_{i}^t \in \{0,1\}$.
Additionally, define $B_i = \{(t_0, t_1) | s_i^{(t_0 - 1)} = 0, s_i^{(t_1 + 1)} = 0, \text{ and } s_i^{j} = 1, t_0 \leq j \leq t_1\}$ as the set of 'on-blocks' for each generator and $D_i$ (defined similarly) be the set of a generator's `off-blocks`. 
There are a few system-level variables, as well: the system-level net load (demand - solar - wind) is $L_t$, and the required ramping between hours is $R_t = L_t - L_{t-1}$. 
Finally, the generation capacity is calculated as $P_t = \sum_{g_i \in G} s_i^t * \overline{p_i}$ 
and ramping capacity as $D_t = \sum_{g_i \in G} s_i^t * r_i$. 

\begin{comment}
\begin{algorithm}
\caption{ Min Up/Down Constraints}
\footnotesize
\begin{algorithmic}[1]
\For{$g_i \in G$}
    \For{$(t_0, t_1) \in B_i$}
        \If{$t_1 - t_0 < mu_i$}
            \State $s_i^{t_0 : t_0 + mu_i} = 1$
        \EndIf
    \EndFor
    \For{$(t_0, t_1) \in C_i$}
        \If{$t_1 - t_0 < md_i$}
            \State $s_i^{t_0 : t_0 + md_i} = 0$
        \EndIf
    \EndFor
\EndFor
\end{algorithmic}
\label{alg:min_up}
\end{algorithm}
\end{comment}

\subsubsection{Surgical repair}
This step ensures that the generator constraints (minimum up and down time) are met and that sufficient generation is online at all times. First, the algorithm iterates through each generator's on-blocks and off-blocks, setting $s_i^t = 1$ until the minimum up time is met and $s_i^t = 0$  until minimum down time is met. Then, the algorithm iterates through each timestep and turns on the cheapest generators until $L_t \leq P_t - \mu$  and $R_t \leq D_t - \mu$ for some threshold value $\mu$. 

\begin{comment}
\begin{algorithm}
\caption{Sufficient and Minimal Generation}
\footnotesize
\begin{algorithmic}[1]
\State G_{sorted} = sort($G$, by = $c_i$)
\For{$t$ in $1:72$}
    \State $i = 1$, $j = 1$
    \While{($L_t > P_t - \mu || R_t > D_t - \mu$)}
        $s_i^t = 1$, $i += 1$
    \EndWhile
    \While{$L_t < P_t + \epsilon$}
        $s_j^t = 0$, $j += 1$
    \EndWhile
\EndFor
\Return $economic\_dispatch(\textbf{s})$
\end{algorithmic}
\label{alg:gen_amt}
\end{algorithm}
\end{comment}

\subsubsection{Economic generator repairs}
This step seeks to reduce the amount of excess online generation at each time step. Taking advantage of the fact that the economic dispatch LP for fixed generator statuses is computationally cheap, the algorithm iterates through each timestep and turns off the most expensive generators until $L_t \geq P_t + \epsilon$ for some threshold value $\epsilon$, ensuring that no generator min-up constraints are violated in the process. It then quickly solves the economic dispatch LP to verify that the system is feasible. The threshold values $\mu$ and $\epsilon$ represent the generation met by the battery.

\subsubsection{Head/tail trim }
As a result of the baseline prediction and the previous steps, generators are often turned on earlier than needed or kept on too long. To accommodate this, algorithm \ref{alg:gen_trim} trims the head and tail of the generator blocks where possible. Taking the result of the economic dispatch run in step $2$ (which gives generator dispatches $p_i^t$), this algorithm iterates through the on-blocks of each generator and calculates its percentage utilization $u_i^{t_0:t_1}$ for the first and last few hours of each on-block. If the utilization is less than a threshold, the generator is turned off for the first and last few hours of each block. 

\begin{algorithm}
\caption{Generation Head/Tail Trim}
\footnotesize
\begin{algorithmic}[1]
\For{$g_i \in G$}
    \For{$(t_0, t_1) \in B_i$}
        \State $u_i^0 = (\sum_{t = t_0}^{t_0 + 3} s_i^t*(p_i^t - \underline{p_i}))/(\overline{p_i} - \underline{p_i})$
        \State $u_i^1 = (\sum_{t = t_1 - 3}^{t_1} s_i^t*(p_i^t - \underline{p_i}))/(\overline{p_i} - \underline{p_i})$
        \If{$u_i^0 < T$}
            \State $s_i^{t_0 : t_0 + 3} = 0$
        \EndIf
        \If{$u_i^1 < T$}
            \State $s_i^{t_1 - 3 : t_1} = 0$
        \EndIf
    \EndFor
\EndFor
\end{algorithmic}
\label{alg:gen_trim}
\end{algorithm}

The utility of these post-processing steps is demonstrated in Figure \ref{fig:predictions}, which compares the predicted generator status before and after the refinements. Even though the baseline prediction in this instance predicts the ground-truth statuses with  98.7\% accuracy, it is still infeasible since it violates the generator minimum-up and minimum-down constraints. The post-processing steps restore feasibility to the sample, and result in a better prediction than relying solely on the solver with 0.25\% MIP gap solution, even without warm-starting and the warm-started solver uses the same MIP gap value.
% than the ground truth label, even without warm-starting.

\begin{figure*}[h!]
    \centering
    \includegraphics[width=0.75\linewidth,height=6cm]{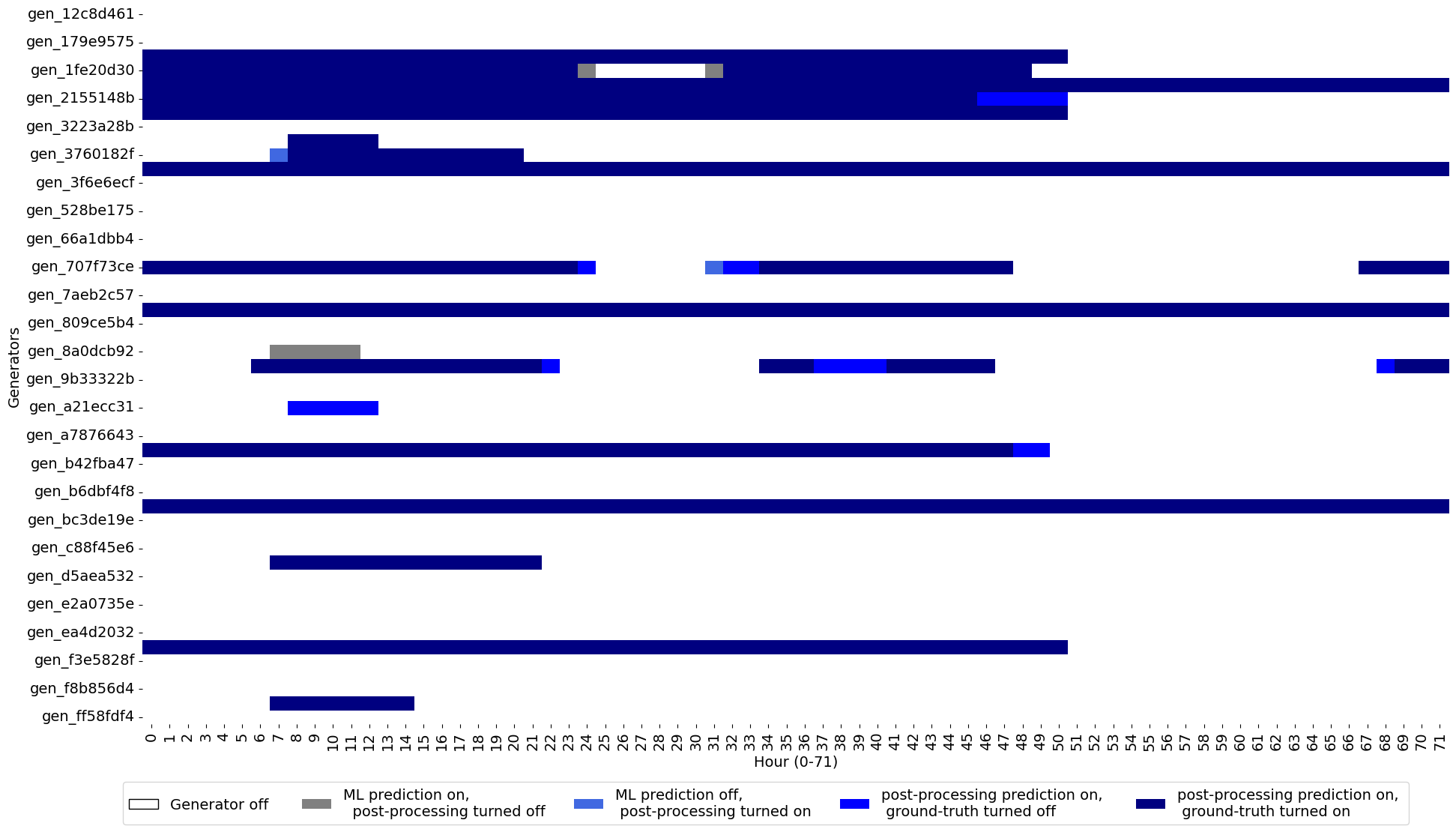}
    \caption{Predicted generator statuses on a sample data point where the model prediction obtained a lower cost than the ground truth.  Grey values indicate the generator was turned off by the post-processing, while light blue indicates the generator was turned on by post-processing.}
    \label{fig:predictions}
\end{figure*}

\subsection{Warm-Start with Prediction Thresholds}\label{sec:ws}
Solving an MIP is computationally expensive because the disjoint shape of the problem's drives the feasible region to be non-differentiable  %difficult for the solver to navigate the search space. 
Warm-starting the binary decision variables can therefore significantly reduce the computation time by informing the solver where to look. Using the generator status predictions as a warm start will improve both feasibility and provide a solution  at or below the optimality gap \cite{rl_warmstart}. 

The computation time of the warm-started MIP can be further decreased by reducing the number of decision variables. Similar to how \cite{milp_lp_short} predicts the generator statuses for only the most-trusted generators, we fix a generator prediction for any variable that the ML model predicted with a sufficiently high threshold value, thereby reducing the MIP problem size. This technique is referred to as \textit{threshold} in the results section.
% \begin{table}[h!]
%     \centering
%     \begin{tabular}{|c|c|c|c|c|c|c|c|c|}
%     \hline
%        label & P.P & W.S & Th.  & \textbf{T.R (\%)} & \textbf{O.R (50)} & \textbf{O.R (25)} & \textbf{O.R (75)} & \textbf{\% Feas.} \\
%        \hline 
%       M1 &  &  & & 0.3 &  1.004 & 1.0019 & 1.0069 & 31.6 \\
%       M2 & \checkmark &  &  & 1.5 &  1.0025 &1.0012 & 1.0041 &  100 \\
%        M3 &  & \checkmark &  & 61.2 &  1.002 & 1.0008 & 1.0035 & 77.6 \\
%        M4 &  & \checkmark & \checkmark & 14.4 &  1.0008 & 1.000 & 1.0018 & 98.6 \\
%        M5 & \checkmark & \checkmark &  & 52.6 &  1.002 & 1.0008 & 1.0037 & 100 \\
%        M6 & \checkmark & \checkmark & \checkmark & 13.4 &  \textbf{1.0007}& \textbf{1.000} & \textbf{1.0018} & \textbf{100} \\
%        \hline 
%      % Gurobi & M1 &  &  &   & 0.4 & 1.004 & 31.6 \\
%      % Gurobi  & M2 & \checkmark &  &   & 1.5 & 1.0025 & 100 \\
%      %  Gurobi  & M3 &  & \checkmark & \checkmark  & 0.5 & 1.004 & 31.6 \\
%      %  \textbf{Gurobi}  & M4 & \checkmark & \checkmark & \checkmark  & 1.7 & \textbf{1.0025} & \textbf{100} \\
%        \hline 
%     \end{tabular}
%     \caption{Various model iterations for each solver. \textit{P.P} = post-processing, \textit{W.S} = warm-start, \textit{Th.} = threshold, \textit{T.R} = time ratio, \textit{O.R} = optimality ratio}
%     \label{tab:best_model}
% \end{table}

\section{Implementation \& Results}\label{sec: res}

This section presents the implementation details, conducted experiments and their results.

\subsection{Implementation}

\subsubsection{Data generation}\label{sec:data}

The competition provided a dataset of $2620$ instances. In addition to generator-specific characteristics that remained constant across all samples, each instance consisted of hourly load, solar, and wind forecasts over 72 hours and the resulting hourly generator status for the 51 thermal generators. The MIP was solved using Gurobi with a MIP gap of 0.25\% to match the competitions dataset MIP gap and to accelerate the generation of training samples.

To increase the size and scope of the dataset, we utilized the provided instances to generate additional data by perturbing the solar, wind and load values. We applied simple perturbations on the profiles to include a diverse set of potential operational conditions such as zero net load and 100\% renewable generation of the load. This step created more samples for the ML model training, and  improved the model performance on challenging scenarios. The final dataset consisted of over 420,000 diverse samples. The initial competition-provided dataset is used to validate the models in this paper.

\subsubsection{Model and Computation}
Deep learning models, data generation pipelines, and post-processing heuristics were implemented in Python 3.12. The neural networks were built and trained using the PyTorch framework. Model training was executed on a computing cluster utilizing two NVIDIA A100 GPUs. To accelerate training, we distributed the workload using PyTorch Distributed Data Parallel (DDP) with the NCCL backend, employing a batch size of $B = 1024$ and gradient clipping of $\gamma = 1.0$.

For inference and downstream optimization, the pipeline was executed locally on an Apple M4 machine equipped with 16 GB of unified memory. The MILP formulations were modeled using Scipy package and optimized using HiGHS solver via its Python APIs. Finally, the binary decision threshold $\tau$ was empirically tuned during this inference phase.

\subsection{Validation}

\subsubsection{Proposed Pipeline Validation}
The model (and variations) introduced in this paper were validated using $500$ samples provided by the competition \cite{epri}.
The evaluated models were compared against a baseline approach that used the HiGHS solver independently. Performance was measured using three key metrics: Time Ratio (T.R.), calculated by dividing the model's solve time by the HiGHS solve time; Optimality Ratio (O.R.), defined as the operational cost achieved by the model divided by the cost obtained from the HiGHS solver (reported in quartiles); and finally, Feasibility, which represents the percentage of test set instances that yielded physically feasible solutions.

We present here the results of the ML model. We explored all possible variations of the pipeline, with regard of enabling or disabling the stages mentioned in Section II. In the iterations that use warm-starting, the MIP was solved using  HiGHS solver with a MIP gap of $0.25\%$, both with and without the fixed-variable threshold. The output of these models were then fed into an LP solver to obtain the dispath of each generators. 

The configruration and the performance for each model iteration are presented in Table \ref{tab:best_model}, with the highest performing models bolded. Models were scored based on the ratio between the LP cost from the predicted generator status and the ground truth LP cost over all $500$ samples. We also analyze the computation time ratio between the ML model and the full MILP computation solve time. The six model iterations are labeled $M1 - M6$, and are referenced as such throughout the remainder of this section. 

\begin{table}[h!]
    \centering
    
    % --- TOP BLOCK: Model Configurations ---
    \begin{tabular}{cccccccc}
    \hline
       \textbf{label} & \textbf{P.P} & \textbf{W.S} & \textbf{Th.} & \textbf{label} & \textbf{P.P} & \textbf{W.S} & \textbf{Th. }\\
       \hline 
      M1 & & & & M2 & \checkmark & & \\
      M3 & & \checkmark & & M4 & & \checkmark & \checkmark \\
      M5 & \checkmark & \checkmark & & M6 & \checkmark & \checkmark & \checkmark \\
       \hline 
    \end{tabular}
    
    \vspace{1em} % Adds a little breathing room
    \hrule % The requested horizontal line
    \vspace{1em}
    
    % --- BOTTOM BLOCK: Performance Metrics ---
    \begin{tabular}{cccccc}
    \hline
       \textbf{label} & \textbf{T.R (\%)} & \textbf{O.R (25)} & \textbf{O.R (50)} & \textbf{O.R (75)} & \textbf{\% Feas.} \\
       \hline 
      M1 & \textbf{0.03} & 1.0019 & 1.004 & 1.0069 & 31.6 \\
      M2 & .16 & 1.0012 & 1.0025 & 1.0041 & 100 \\
       M3 & 5.80 & 1.0008 & 1.002 & 1.0035 & 77.6 \\
       M4 & 1.56 & 1.0000 & 1.0008 & 1.0018 & 98.6 \\
       M5 & 5.34 & 1.0008 & 1.002 & 1.0037 & 100 \\
       \textbf{M6} & 1.46 & \textbf{1.0000} & \textbf{1.0007} & \textbf{1.0018} & \textbf{100} \\
       \hline 
    \end{tabular}
    
    \caption{Various model iterations for each solver. \textit{P.P} = post-processing, \textit{W.S} = warm-start, \textit{Th.} = threshold, \textit{T.R} = time ratio, \textit{O.R} = optimality ratio.}
    \label{tab:best_model}
\end{table}

\begin{comment}
\begin{table}[h!]
    \centering
    
    % --- TOP BLOCK: Model Configurations ---
    \begin{tabular}{|c|c|c|c|}
    \hline
       label & P.P & W.S & Th. \\
       \hline 
      M1 &  &  & \\
      M2 & \checkmark &  & \\
       M3 &  & \checkmark & \\
       M4 &  & \checkmark & \checkmark \\
       M5 & \checkmark & \checkmark & \\
       M6 & \checkmark & \checkmark & \checkmark \\
       \hline 
    \end{tabular}
    
    \vspace{1em} % Adds a little breathing room
    \hrule % The requested horizontal line
    \vspace{1em}
    
    % --- BOTTOM BLOCK: Performance Metrics ---
    \begin{tabular}{|c|c|c|c|c|c|}
    \hline
       label & \textbf{T.R (\%)} & \textbf{O.R (25)} & \textbf{O.R (50)} & \textbf{O.R (75)} & \textbf{\% Feas.} \\
       \hline 
      M1 & 0.03 & 1.0019 & 1.004 & 1.0069 & 31.6 \\
      M2 & .16 & 1.0012 & 1.0025 & 1.0041 & 100 \\
       M3 & 5.80 & 1.0008 & 1.002 & 1.0035 & 77.6 \\
       M4 & 1.56 & 1.0000 & 1.0008 & 1.0018 & 98.6 \\
       M5 & 5.34 & 1.0008 & 1.002 & 1.0037 & 100 \\
       \textbf{M6} & 1.46 & \textbf{1.0000} & \textbf{1.0007} & \textbf{1.0018} & \textbf{100} \\
       \hline 
    \end{tabular}
    
    \caption{Various model iterations for each solver. \textit{P.P} = post-processing, \textit{W.S} = warm-start, \textit{Th.} = threshold, \textit{T.R} = time ratio, \textit{O.R} = optimality ratio.}
    \label{tab:best_model}
\end{table}
\end{comment}

As shown in Table \ref{tab:best_model}. It is worth mentioning that while M1 model (baseline transformer) obtained a 98\% accuracy when evaluated as commitment status, the predictions resulted in a median LP cost increase of 0.4\% with only 31.6\% feasible samples, which highlights the need for post-processing steps. 
The heuristic-based post processing (M2) significantly increased the feasibility ratio to 100\%, with enhanced optimality ratio compared to M1. M3 and M4 variants were the heuristic-based post processing was disabled, but the deep learning prediction were fed as warm start to the MIP solver (with fixed variables in M4). The cost ratio quartiles were enhanced, but with reduced feasibility ratios. Variant M5, which had the post-processing and warm start for a solver retained 100\% feasibility but with higher optimality ratio. Finally, variant M6 were all the stages of the pipeline were enabled, achieved the best overall feasibility ratio and optimality ratio. It is evident from the table that using the deep learning predictions as warm start for a solver enhances the cost ratio, on the cost of speed, but fixing the variables with respect to a threshold helps in speeding up the process. 
 %vthe highest performing model (M6) produced 100\% feasibility with a median cost increase of 0.07\%. The ML models produce generator status predictions in $0.03 - 5.8\%$ of the time taken by the MILP solver. Implementing the threshold values with the warm-start (i.e moving from M3 to M4 or M5 to M6) significantly reduces the size of the problem, thus decreasing the time ratio.

\begin{figure}[h!]
    \centering

    % First figure
    \begin{subfigure}[h!]{0.95\linewidth}
        \centering
        \includegraphics[width=\linewidth]{Figures/cdf_6models.png}
    \end{subfigure}

    \caption{Distribution of optimality ratios for models M1 - M6, capped at 1.01. Only feasible samples are considered. }
    \label{fig:trans_dist}
\end{figure}
Figure \ref{fig:trans_dist} shows the distribution of optimality ratios over all models, considering only the samples that had feasible predictions with all models. The M1 model has the lowest slope, with a maximum optimality ratio of 2.2.  Implementing either the post-processing steps or the warm-start (M2- M4) increases the slope slightly, while models M5 and M6 have the lowest scoring optimality ratios. One of the most notable results from this paper is that around 20\% of all samples predicted with the M6 model have an optimality ratio below 1.  
This indicates that the M6 model predicted a solution \textit{better} than the Highs solver, even though both were solved with a MIP gap of 0.25\%.

Finally, we compare the impacts of each individual refinement stage. Figure \ref{fig:three_results} shows the calculated main effects on feasibility, optimality, and computation time for the post-processing steps,  warm-starting the algorithm, and  warm-starting with a threshold. All factors independently decrease in the optimality ratio, although warm-starting has a more significant impact than including the post-processing heuristics (indicated by a steeper slope). Similarly, all factors independently increase the feasibility, but warm-starting the solver only increases the feasibility to 98.8\%, while including the post-processing steps creates feasibility for all samples. Finally, warm-starting the algorithm significantly increases the computation time, but this impact is mitigated with fixed-variable thresholding. 
% \begin{figure}
%     \centering
%     \includegraphics[width=0.75\linewidth]{Figures/optimality_ratio_dist.png}
%     \caption{Distribution of optimality ratios for the M1 model versus the M4 model, capped at 1.02.}
%     \label{fig:trans_dist}
% \end{figure}

% The two heatmaps show the interaction terms between each factor. As shown in Subfigure \ref{fig:feas_hm}, including the post-processing steps guarantees feasibility, regardless of the other two factors. Without the post processing, warm-starting without thresholds obtained feasibility on only 87.9\% of samples, and on 98.8\% of samples with thresholds. The impact of the post-processing steps on the optimality ratio is less significant;  the lowest ratio can be obtained without the post-processing steps. Including the threshold values in the warm-start lowers the optimality ratio from 1.0014 to 1.0008. 
\begin{figure}[h!]
    \centering
    % % First Subfigure
    % \begin{subfigure}[b]{0.3\linewidth}
    %     \centering
    %     \includegraphics[width=\linewidth]{Figures/highs_feas_hm.png}
    %     \caption{Feasibility heat map.}
    %     \label{fig:feas_hm}
    % \end{subfigure}
    % \hfill
    % Second Subfigure
    \begin{subfigure}[h]{\linewidth}
        \centering
        \includegraphics[width=0.75\linewidth,%height=4cm
        ]{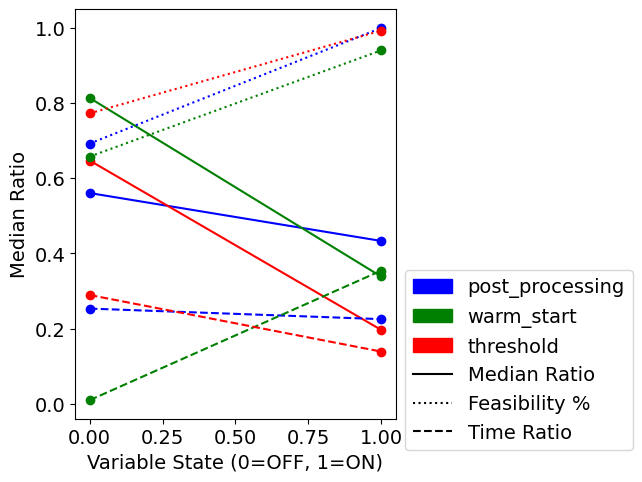}
        % \caption{HiGHS.}
        \label{fig:main_effect}
    \end{subfigure}
    % \hfill
    % \begin{subfigure}[h]{0.45\linewidth}
    %     \centering
    %     \includegraphics[width=\linewidth]{Figures/grb_me.png}
    %     \caption{Gurobi.}
    % \end{subfigure}
    % % Third Subfigure
    % \begin{subfigure}[b]{0.3\linewidth}
    %     \centering
    %     \includegraphics[width=\linewidth]{Figures/highs_opt_hm.png}
    %     \caption{Optimality heat map.}
    %     \label{fig:opt_hm}
    % \end{subfigure}
    
    \caption{Main effects (scaled from 0 to 1) for optimality ratio and feasibility.}
    \label{fig:three_results}
\end{figure}

\subsubsection{Comparison against Other Methods}

To analyze the effect of the baseline transformer architecture on the overall predictions, a neural network (NN) (used in \cite{feas_restore} and Long-short-term-memory network (LSTM) (used in \cite{ramesh2023spatio} was trained on the same data, then augmented with all three heuristic steps.  The results of these models are displayed in Table \ref{tab:ml_model}. The transformer M6 model generated the predictions with the lowest cost ratio, but took longer on average to run; the LSTM and the NN have a computation time ratio similar to the M1 and M2 model, even though they are warm-started. This suggests that while the smaller models do provide better computational speed-ups, the transformer provides a better initial prediction that leads to a more optimal solution when paired with the post-processing and warm-starting.

\begin{table}[h!]
    \centering
    \begin{tabular}{cccccc}
    \hline
       \textbf{Model}  & \textbf{T.R (\%)} &\textbf{O.R. (25)} & \textbf{O.R. (50)} & \textbf{O.R. (75)} &\% \textbf{Feas.}  \\
       \hline 
       M6 & 1.46 &\textbf{1.0000}& \textbf{1.0007}&\textbf{1.0018}
       & 100 \\
       NN & \textbf{0.05} & 1.0036 & 1.0057& 1.0089 & 100 \\
       LSTM  & 0.39& 1.0019 & 1.0039 & 1.0065 & 99.6 \\ 
       \hline
    \end{tabular}
    \caption{Comparison of ML model baseline. }
    \label{tab:ml_model}
\end{table}

It is worth mentioning that the proposed transformer model is slower than other ML models. Because transformer models are inherently bigger and slower than LSTM and NN architectures. However, The significant difference in speed is due to the fact that the transformer architecture inference is only GPU-optimized, and these numbers are reported on Apple M4 chip, which gives LSTM and NN speed advantages.

\section{Conclusion}\label{sec: conc}
% Unit commitment is  a well-studied problem in power systems operations, due to its high complexity. This paper introduces a machine-learning model designed to enhance a unit commitment solver to produce feasible and near-optimal solutions in a fraction of the computation time. Starting with a baseline Transformer that achieves high accuracy on the ground truth generator predictions, this work introduces additional heuristics that ensure generator constraint satisfaction, as well as implementing a probability-threshold strategy to dynamically reduce the size of the MIP and improve the computation time of the warm-started solver. The final model introduced in this paper produced 100\% feasible samples with a median cost increase of 0.07\%. Importantly, the final model found predictions that improved on the cost of the ground-truth data, out-performing the solver in 10\% of the computation time. 
This paper presents a machine learning based pipeline for 3 day-ahead hourly unit commitment. The proposed methods consists of three stages; the first stage is a self-attention model, followed by a post-processing heuristics step, which outputs are fed as a warm start for a UC MILP solver. The proposed method was validated on a single bus system that represents the Irish system, and compared against other machine learning methods. The results of the experiments clearly demonstrated the significant computation speed gained from using machine learning methods as a warm-start for UC solvers. 

There are a few notable limitations to this paper that should be acknowledged. As mentioned, the demand, wind, and solar are assumed to be perfect forecasts. Including forecast uncertainty would require adjustments to guarantee feasibility. Furthermore, this work is based on a copper-plate model and does not include transmission constraints, so conclusions cannot be drawn as to how successful this proposed pipeline would be on a more complex network. 
Future work should apply this framework to various larger power systems networks and incorporate more accurate data and constraints.

\section*{Acknowledgment}
We would like to thank EPRI for organizing and hosting the `AI-ccelerating Unit Commitment' competition, as well as providing the dataset and problem formulation that this work was founded on \cite{epri}.

\bibliographystyle{IEEEtran}
\bibliography{references}

\end{document}